\documentclass[11pt]{article}

\usepackage{amsmath,amssymb,latexsym}
\usepackage[T1]{fontenc}
\usepackage{epsfig}
\usepackage{fullpage}
\usepackage{graphicx}
\usepackage{makeidx}
\usepackage{slashed}
\usepackage{color}
\usepackage[normalem]{ulem}
\usepackage{subcaption}
\usepackage{feynmp}
\usepackage{cite}
\usepackage{hyperref}

\long\def\symbolfootnote[#1]#2{\begingroup%
\def\thefootnote{\fnsymbol{footnote}}\footnote[#1]{#2}\endgroup}

\newcommand \slsh [1] {\not\!{#1}}

\begin{document}


\vspace*{0.5cm}

\begin{center}
  {\Large\sc Next-to-soft radiation from a different
    angle}\\[10ex]
  {Melissa van Beekveld$^a$\footnote{melissa.vanbeekveld@physics.ox.ac.uk}, Abhinava Danish$^b$\footnote{abhinavadan@gmail.com}, Eric
    Laenen$^{c,d,e}$\footnote{eric.laenen@nikhef.nl},\\
    Sourav Pal$^f$\footnote{sourav@prl.res.in}, Anurag Tripathi$^b$\footnote{tripathi@phy.iith.ac.in} and Chris
    D. White$^{g}$\footnote{christopher.white@qmul.ac.uk}}\\[1cm]

{\it
  $^a${Rudolf Peierls Centre for Theoretical Physics, Clarendon Laboratory,\\ Parks Road, University of Oxford, Oxford OX1 3PU, United Kingdom}\\
  $^b${Department of Physics, Indian Institute of Technology Hyderabad, Kandi,\\ Sangareddy, Telangana State 502285, India}\\
  $^c${Institute of Physics, University of Amsterdam, Science Park 904,\\ 1098 XH Amsterdam, The Netherlands}\\
  $^d${Nikhef, Theory Group, Science Park 105, 1098 XG Amsterdam,\\ The Netherlands}\\
  $^e${Institute for Theoretical Physics, Utrecht University, Leuvenlaan 4,\\ 3584 CE Utrecht, The Netherlands}\\
  $^f${Theoretical Physics Division, Physical Research Laboratory, \\ Navrangpura, Ahmedabad 380009, India}\\
  $^g${Centre for Theoretical Physics, Department of
    Physics and Astronomy,\\ Queen Mary University of London, Mile
    End Road, London E1 4NS, United Kingdom}\\}
\end{center}

\vspace{1.5cm}

\begin{abstract}
\noindent{} Soft and collinear radiation in collider processes can be
described in a universal way, that is independent of the underlying
process. Recent years have seen a number of approaches for probing
whether radiation beyond the leading soft approximation can also be
systematically classified. In this paper, we study a formula that
captures the leading next-to-soft QCD radiation affecting processes
with both final- and initial-state partons, by shifting the momenta in
the non-radiative squared amplitude. We first examine W+jet
production, and show that a previously derived formula of this type
indeed holds in the case in which massive colour singlet particles are
present in the final state. Next, we develop a physical understanding
of the momentum shifts, showing precisely how they disrupt the
well-known angular ordering property of leading soft radiation.
\end{abstract}

\newpage
\reversemarginpar

\section{Introduction}

The calculation of scattering amplitudes in perturbative quantum field
theory continues to be an area of intense activity, due to its many
applications to current and future collider experiments. Whilst it is
often possible to obtain complete amplitudes at a given order in the
coupling constant, we sometimes wish to consider approximate results,
particularly where these can be resummed to all orders in perturbation
theory. A particularly well-studied case is the emission of soft and /
or collinear radiation dressing an underlying scattering
amplitude. This generates infrared singularities, which will cancel
for suitably inclusive observables, such as total hadronic
cross-sections. However, large contributions remain in perturbation
theory, typically involving large logarithms of dimensionless energy
ratios. A variety of methods have been developed for resumming such
contributions (see
e.g. refs.~\cite{Parisi:1979xd,Curci:1979am,Sterman:1986aj,Catani:1989ne,Gatheral:1983cz,Frenkel:1984pz,Sterman:1981jc,Korchemsky:1992xv,Korchemsky:1993uz,Becher:2006nr,Schwartz:2007ib,Bauer:2008dt,Chiu:2009mg}),
all of which rely on the tight relationship between kinematically
enhanced terms and infrared singularities, plus the fact that soft and
collinear factorisation can be described in terms of universal
functions acting on arbitrary amplitudes. The latter property has a
simple quantum mechanical interpretation: soft radiation has zero
momentum, and thus an infinite Compton wavelength by the uncertainty
principle. Thus, it cannot resolve the details of the underlying
scattering amplitude that produced the hard outgoing particles. A
similar story applies to collinear radiation, which instead has a
zero transverse momentum relative to a given outgoing particle.

Heuristic arguments such as these are also useful for understanding the
wider implications of soft radiation. Crucial for this paper will be a
particular property of soft radiation that is emitted from pairs of
(colour) charges, or {\it dipoles}, in QED or QCD. Including all
possible quantum interference contributions in the squared amplitude,
one finds in QED that radiation is confined to a cone around each
charged particle, whose half-angle coincides with the angle between
the two charged particle momenta. This is known as the {\it Chudakov
  effect}, and textbook treatments may be found in
refs.~\cite{Ellis:1991qj,Campbell:2017hsr}. A corresponding effect
holds in QCD, where for two colour charges there is no radiation (at
leading soft level) outside the cones surrounding each particle. For
more complicated configurations of partons, clusters of particles
radiate according to their combined colour charge at sufficiently
large angles. All of these phenomena have a common quantum mechanical
origin similar to that already mentioned above: at large angles, the
wavelength of the emitted radiation is such that it can only notice
the combined colour charge of a given subset of partons. If this
combined charge happens to be zero (or colour singlet in the QCD
case), then there is no radiation at large angles.

So much for soft radiation, whose properties are already
well-known. Until recently, much less has been known about how to
systematically classify the properties of radiation at subleading
order in a systematic expansion in the total radiated momentum. The
frontier of such attempts is at next-to-leading power (NLP), and the
last few years have seen an increasing number of techniques aimed at
clarifying whether any universal statements can be made about such
radiation, including its possible resummation. The range of
methods~\cite{Grunberg:2009yi,Soar:2009yh,
  Moch:2009hr,Moch:2009mu,Laenen:2010uz,Laenen:2008gt,deFlorian:2014vta,Presti:2014lqa,Bonocore:2015esa,Bonocore:2016awd,Bonocore:2020xuj,Gervais:2017yxv,Gervais:2017zky,Gervais:2017zdb,Laenen:2020nrt,DelDuca:2017twk,vanBeekveld:2019prq,Bonocore:2014wua,Bahjat-Abbas:2018hpv,Ebert:2018lzn,Boughezal:2018mvf,Boughezal:2019ggi,Bahjat-Abbas:2019fqa,Ajjath:2020ulr,Ajjath:2020sjk,Ajjath:2020lwb,Ahmed:2020caw,Ahmed:2020nci,Ajjath:2021lvg,Kolodrubetz:2016uim,Moult:2016fqy,Feige:2017zci,Beneke:2017ztn,Beneke:2018rbh,Bhattacharya:2018vph,Beneke:2019kgv,Bodwin:2021epw,Moult:2019mog,Beneke:2019oqx,Liu:2019oav,Liu:2020tzd,Boughezal:2016zws,Moult:2017rpl,Chang:2017atu,Moult:2018jjd,Beneke:2018gvs,Ebert:2018gsn,Beneke:2019mua,Moult:2019uhz,Liu:2020ydl,Liu:2020eqe,Wang:2019mym,Beneke:2020ibj,vanBeekveld:2021mxn}
-- some of them inspired by the much earlier work of
refs.~\cite{Low:1958sn,Burnett:1967km,DelDuca:1990gz} -- mirrors that
used for soft radiation, and this body of work is ultimately motivated
by the fact that the numerical impact of such contributions may be
needed to increase the theoretical precision of collider physics
observables~\cite{Kramer:1996iq, Ball:2013bra, Bonvini:2014qga,
  Anastasiou:2015ema, Anastasiou:2016cez, vanBeekveld:2019cks,
  vanBeekveld:2021hhv, Ajjath:2021lvg}. As well as studies aiming to
develop new resummation formulae, there is also scope for case studies
that look at well-defined consequences of next-to-soft radiation, in
order to build up our collective intuition of how it behaves. The aim
of this paper is to carry out such a case study.

Our starting point is to consider a formula -- first derived in
ref.~\cite{DelDuca:2017twk} and extended in
ref~\cite{vanBeekveld:2019prq} -- that states that leading
next-to-soft radiative contributions can be expressed in a
particularly compact and elegant form. That is, the squared amplitude
including such radiation can be written in terms of non-radiative
amplitudes, but where distinct pairs of partonic momenta are shifted
in a prescribed way. The shifted squared amplitudes are then dressed
by overall factors which are identical to those that occur in the
leading soft limit. Potential uses of such formulae include increasing
the precision of numerical NLO calculations, and similar comments,
independently and using different methods, have been made in
refs.~\cite{Boughezal:2016zws,Boughezal:2018mvf,Boughezal:2019ggi}. However,
the obvious similarity of these momentum-shift formula to their
leading-soft counterparts means they are an excellent starting point
for examining the physics of next-to-soft radiation in a particularly
transparent way.

In this paper, we will first review the momentum-shift formulae of
refs.~\cite{DelDuca:2017twk,vanBeekveld:2019prq}, and introduce them
by considering a process that has not previously been considered
before in this approach. That is, we will consider radiative
corrections to $W$ production in association with an additional hard
jet\footnote{For previous work on this process in the context of
next-to-soft physics, see
refs.~\cite{Boughezal:2019ggi,Sterman:2022lki}.}. This is more general
than either of the processes considered previously in this
approach. Reference~\cite{DelDuca:2017twk} looked only at
colour-singlet final states, whereas ref.~\cite{vanBeekveld:2019prq}
considered only final states with massless particles. As we will yet
again see, leading next-to-soft radiative corrections take the form of
a series of dipole-like terms~\footnote{Very recently,
ref.~\cite{Czakon:2023tld} has provided alternative formulae for
capturing next-to-soft radiation, which also work at loop level.}.

Next, we consider the effect of the momentum-shift formulae on the
emission of soft radiation from a single final-state dipole. We will
briefly review the well-known calculation of how soft gluon emission
is confined to cones surrounding each hard particle, before correcting
this to include the effects of the momentum-shifts, and hence leading
next-to-soft effects. We will show explicitly that the next-to-soft
corrections break the angular ordering property, in that they lead to
emission outside of the usual angular region. That this property is
not preserved beyond leading soft level will perhaps not surprise
anyone. However, the mechanism by which this happens, including the
details of the calculation, are interesting. Furthermore, given that
the origin of the momentum shifts is well-understood as arising from
orbital angular momentum
effects~\cite{DelDuca:2017twk,vanBeekveld:2019prq}, we will be able to
precisely interpret the physics of how angular ordering breaks
down. We believe that this story offers novel insights into the
physics of next-to-soft radiation that, as well as being compelling in
themselves, may be of broader use. 

The structure of our paper is as follows. In section~\ref{sec:Wg}, we
examine $W$ + jet production at NLO, showing that the inclusion of
radiative corrections up to next-to-soft level reproduces the same
momentum-shift formula as was found for prompt production in
ref.~\cite{vanBeekveld:2019prq}. After drawing attention to the
dipole-like nature of this formula, in section~\ref{sec:ordering} we
show that next-to-soft corrections lead to radiation outside the cone
regions associated with leading soft radiation, and intepret the
physics of this effect in detail. In section~\ref{sec:conclude}, we
discuss our results and conclude.

\section{A momentum-shift formula for $W$ plus jet production}
\label{sec:Wg}

\subsection{$W$ plus jet production up to NLO}
\label{sec:NLO}

\begin{figure}
  \begin{center}
    \scalebox{0.7}{\includegraphics{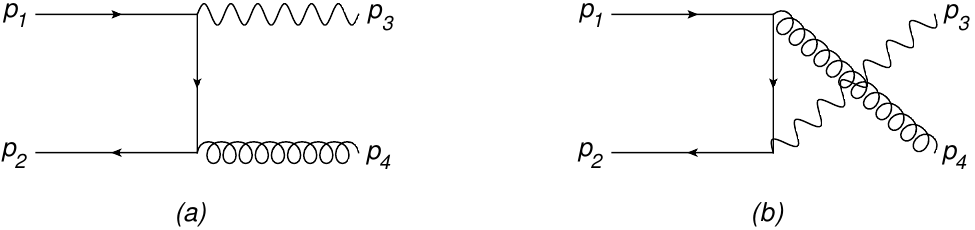}}
    \caption{Leading order diagrams for $W$ plus jet production in the
      $q\bar{q}$ channel.}
    \label{fig:LOdiags}
  \end{center}
\end{figure}

We start by considering the LO process
\begin{equation}
  q(p_1)+\bar{q}(p_2)\rightarrow W(p_3)+g(p_4),
  \label{LOprocess}
\end{equation} 
whose Feynman diagrams are shown in figure~\ref{fig:LOdiags}. This is
itself a correction to the Drell-Yan production of a $W$ boson, but we
will consider that the final-state gluon is constrained to be hard
(e.g. through a non-zero transverse momentum requirement), such that
no infrared singularities are present. Our aim is to show how
next-to-soft corrections to this process can be written according to a
certain formula, and we will be able to illustrate our point without
having to consider the alternative partonic channel $qg\rightarrow
W+g$, which in any case can be obtained from crossing. Denoting the
$W$ boson mass by $m$, the various momenta satisfy
\begin{equation}
  p_1^2=p_2^2=p_4^2=0,\quad p_3^2=m^2,
  \label{momenta}
\end{equation}
and we also define the Mandelstam invariants
\begin{equation}
  s=(p_1+p_2)^2,\quad t=(p_1-p_3)^2,\quad u=(p_2-p_3)^2.
  \label{mandies}
\end{equation}
It is also conventional to define the alternative invariants
\begin{equation}
  t_1=t-m^2, \quad u_1=u-m^2,
  \label{t1u1}
\end{equation}
which obey
\begin{equation}
  s+t_1+u_1+m^2=0
\label{momcon}  
\end{equation}
as a consequence of momentum conservation. With this notation, the
squared LO amplitude, summed (averaged) over final (initial) colours
and spins, is given by
\begin{equation}
  \overline{|{\cal M}^{(0)}|}^2=g_s^2 g_w^2\frac{C_F}{N_c}
  \frac{(2m^4-2m^2(t+u)+t^2+u^2)}{2tu},
  \label{LOamp2}
\end{equation}
where $g_s$ and $g_w$ are the strong and electroweak coupling
constants respectively, $C_F$ the quadratic Casimir in the fundamental
representation, and $N_c$ the number of colours.

\begin{figure}
  \begin{center}
    \scalebox{1.0}{\includegraphics{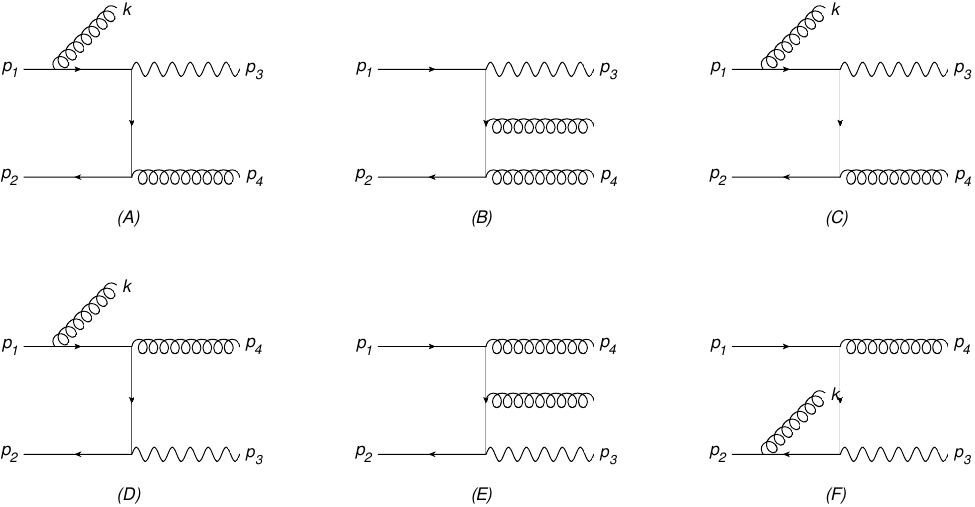}}
    \caption{NLO diagrams to $W$ plus jet production, in which a gluon is radiated off an (anti-)quark.}
    \label{fig:NLOdiags}
  \end{center}
\end{figure}

\begin{figure}
  \begin{center}
    \scalebox{0.7}{\includegraphics{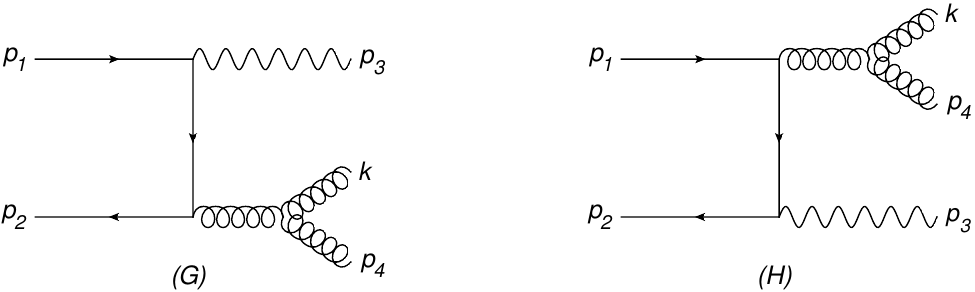}}
    \caption{NLO diagrams to $W$ plus jet production, in which a gluon is radiated off the hard gluon.}
    \label{fig:NLOdiags2}
  \end{center}
\end{figure}

Let us now consider the radiation of an additional gluon, for which
there are two types of diagram. First, there is radiation of a quark
or antiquark, as shown in figure~\ref{fig:NLOdiags}. These diagrams
would also be present in the case of $W\gamma$ production, which was
first calculated at NLO in ref.~\cite{Smith:1989xz}. Next, there are
diagrams in which the gluon is radiated off the final state hard
gluon, as in figure~\ref{fig:NLOdiags2}.
Although the full set of NLO diagrams for $W$+jet production
(including all partonic channels) has been calculated
before~\cite{Giele:1993dj}, full analytic expressions are rarely
reported due to their cumbersome nature. Thus, we have recalculated
these diagrams independently in FORM~\cite{Vermaseren:2000nd} and
FeynCalc~\cite{Kublbeck:1992mt}, finding agreement. Here, we will
report analytic results for the squared and summed / averaged matrix
element expanded to first subleading order in the emitted gluon
momentum. To do so, we can introduce the Mandelstam invariants
\begin{align}
  s&=(p_1+p_2)^2,\quad q_1=(p_1-p_3)^2,\quad q_2=(p_2-p_4)^2,\notag\\
  s_2&=(p_3+p_4)^2,\quad \hat{q}_1=(p_1-p_4)^2,\quad
  \hat{q}_2=(p_2-p_3)^2,
  \label{mandiesNLO}
\end{align}
and
\begin{equation}
  t_k=(p_1-k)^2,\quad u_k=(p_2-k)^2,\quad w_1=(p_3+k)^2,\quad
  w_2=(p_4+k)^2,
  \label{mandiesNLO2}
\end{equation}
where we have adopted notation for ease of comparison with
ref.~\cite{DelDuca:2017twk} (see also ref.~\cite{Frixione:1993yp}).
The various Mandelstam invariants in eqs.~(\ref{mandiesNLO},
\ref{mandiesNLO2}) can be expressed in terms of five independent
invariants, using the relations
\begin{align}
  \hat{q}_1&=m^2-s-t_k-q_1;\notag\\
  s_2&=s+t_k+u_k;\notag\\
  \hat{q}_2&=m^2-s-u_k-q_2;\notag\\
  w_1&=m^2-q_1-q_2-t_k;\notag\\
  w_2&=q_1-q_2-u_k.
  \label{mandyrels}
\end{align}
Next, we can perform the next-to-soft expansion by introducing a
book-keeping parameter $\lambda$ via
\begin{equation}
  t_k\rightarrow \lambda t_k,\quad u_k\rightarrow \lambda u_k,
  \label{tkuk}
\end{equation}
before performing a Laurent expansion in $\lambda$ to first subleading
order. Finally, one sets $\lambda=1$. Compact results are then
obtained upon using a particular Lorentz frame for the final-state
momenta. Following the case of $W\gamma$ production in
ref.~\cite{Smith:1989xz}, we can choose the centre of mass frame of
the $W$ boson and hard gluon, for which an explicit parametrisation is
\begin{align}
  p_1&=(E_1,0,\ldots,0,E_1);\notag\\
  p_2&=(E_2,0,\ldots,0,\omega\sin\psi,\omega\cos\psi-E_1);\notag\\
  k&=(\omega,0,\ldots,0,\omega\sin\psi,\omega\cos\psi);\notag\\
  p_3&=(\omega_W,\ldots,-\omega'\sin\theta_1\sin\theta_2,
  -\omega'\sin\theta_1\cos\theta_2,-\omega'\cos\theta_1);\notag\\
  p_4&=(\omega',\ldots,\omega'\sin\theta_1\sin\theta_2,\omega'
  \sin\theta_1\cos\theta_2,\omega'\cos\theta_1);\notag\\
\end{align}
with
\begin{align}
  E_1&=\frac{s+t_k}{2\sqrt{s_2}},\quad
  E_2=\frac{s+u_k}{2\sqrt{s_2}},\quad
  \cos\psi=\frac{(u_ks-s_2t_k)}{(t_k+u_k)(s+t_k)},\quad\notag\\
  \omega&=-\frac{(t_k+u_k)}{2\sqrt{s_2}},\quad
  \omega'=\frac{(s_2-m^2)}{2\sqrt{s_2}},\quad
  \omega_W=\frac{(s_2+m^2)}{2\sqrt{s}_2}.
\end{align}
Then the (next-to)leading power contributions to the squared matrix
element (summed / averaged over colours and spins) are
\begin{align}
  &  \overline{|{\cal M}^{(1)}|}^2\Big|_{\rm LP}=\frac{1}{N_c}\frac{g_s^4g_w^2}{t_k u_k}
  \Big\{
  -\frac{4 C_F^2\,s\,((\rho^2-1)^2\cos^2\theta_1+(\rho^2+1)^2)}
  {(\rho^2-1)(\cos^2\theta_1-1)}
  \notag\\
&  -\frac{4C_AC_F\,s\,\sqrt{t_k u_k}}{(\rho^2-1)^2(\cos^2\theta_1-1)}
  \frac{((\rho^2-1)^2\cos^2\theta_1+(\rho^2+1)^2)\sin\theta_1\cos\theta_2}{(-2\sin\theta_1
  \cos\theta_2\sqrt{t_k u_k}+\cos\theta_1(t_k-u_k)+t_k+u_k}
  \Big\};\notag\\
  &  \overline{|{\cal M}^{(1)}|}^2\Big|_{\rm NLP}=\frac{1}{N_c}\frac{g_s^4g_w^2}{t_k u_k}
  \Big\{-\frac{16C_F^2}{(\rho^2-1)^3(\cos^2\theta_1-1)^2}\notag\\
  &\times\left[-(\rho-1) (\rho+1) (\rho^4+1) \sin\theta_1 \cos\theta_1
    \cos\theta_2 \sqrt{t_k u_k}+(\rho^2+1) \rho^2 \cos^2\theta_1 (t_k+u_k)
    -(\rho^2+1) \rho^2 (t_k+u_k)
    \right]\notag\\
  &+\frac{2C_FC_A}{(\rho^2-1)^3(\cos^2\theta_1-1)^2(
    -2\sin\theta_1\cos\theta_1\sqrt{t_k u_k}+\cos\theta_1
    (t_k-u_k)+t_k+u_k)^2)}\notag\\
  &\times\Big[
  -(\rho^2 + 
  1) (\sin\theta_1 \cos\theta_2 (\sqrt{t_k u_k}
  (-4 \rho^2 (t_k^2 + 6 t_k u_k + 
  u_k^2) + \rho^4 (-(t_k - u_k)^2)\notag\\
  &+ (t_k - u_k)^2) + 
          4 t_k u_k \sin\theta_1 \cos\theta_2 (\rho^4 t_k + 
             2 \rho^2 (t_k + u_k) - t_k)) - 
       2 (\rho^4 - 4 \rho^2 - 1) t_k u_k (t_k + u_k))  \notag\\
       &  + 2 \cos^2(\theta_1) (4 \sin\theta_1 \cos\theta_2 (t_k u_k \sin\theta_1
       \cos\theta_2 (2 \rho^6 t_k + 
            2 \rho^4 (t_k + u_k) + \rho^2 (t_k + u_k) - t_k + 
            u_k) \notag\\
            &- \rho^2 \sqrt{t_k u_k} (\rho^4 (t_k^2 + u_k^2) + 
            t_k^2 + \rho^2 (t_k + u_k)^2 + 4 t_k u_k + u_k^2)) + 
      t_k u_k ((\rho^6 + \rho^4 + 7 \rho^2 - 
      1) u_k \notag\\
&      - (\rho^6 + \rho^4 - 9 \rho^2 - 
            1) t_k)) + (\rho^2 - 
      1) \cos^4(\theta_1) (-(\rho^2 - 
      1)^2 \sin\theta_1 \cos\theta_2 ((t_k - u_k)^2 \sqrt{t_k u_k}
      \notag\\
      &- 4 t_k^2 u_k \sin\theta_1 \cos\theta_2) - 
      2 t_k u_k (\rho^4 (t_k + u_k) + 2 \rho^2 (u_k - 3 t_k) + t_k + 
      u_k)) \notag\\
      &- (\rho^2 - 
      1)^3 \cos^5(\theta_1) (\sin\theta_1 \cos\theta_2 (t_k - 
      u_k) \sqrt{t_k u_k} (t_k + u_k) - 4 t_k u_k^2)\notag\\
      &+ 4 \rho^2 \cos^3(\theta_1) (2 t_k u_k (\rho^2 (t_k + u_k) + t_k - 
         3 u_k) - (\rho^2 + 1)^2 \sin\theta_1 \cos\theta_2 (t_k - 
         u_k) \sqrt{t_k u_k} (t_k + u_k)) \notag\\
         &+ 
   \cos\theta_1 (\sin\theta_1 \cos\theta_2 (8 (\rho^4 + 
            1) t_k u_k \sin\theta_1 \cos\theta_2 ((2 \rho^2 - 
               1) t_k - 2 (\rho^2 - 
               1) \sin\theta_1 \cos\theta_2 \sqrt{t_k u_k} - 
               u_k) \notag\\
               &- (\rho^2 - 1)^2 (3 \rho^2 + 1) (t_k - u_k) \sqrt{t_k u_k}
               (t_k + u_k)) - 
      4 (\rho^2 + 
      1) t_k u_k (2 \rho^2 t_k + (\rho^4 - 2 \rho^2 - 1) u_k)) \notag\\
      &+ 
   2 (\rho^2 - 1)^3 t_k u_k \cos^6(\theta_1) (t_k - u_k)))\Big]\Big\},
   \label{M1sq}     
\end{align}
where we have introduced the dimensionless parameter
\begin{equation}
  \rho^2=\frac{m^2}{s}.
  \label{rhodef}
\end{equation}
A cross-check of this result can be obtained by considering only the
$C_F^2$ terms, which arise from the diagrams of
figure~\ref{fig:NLOdiags}. These diagrams would also arise in abelian
gauge theory, where $C_F^2$ would be replaced by the appropriate
squared electromagnetic charge of the incoming (anti-)quarks. Then,
one may verify that taking $\rho\rightarrow 0$ (i.e. the limit of zero
$W$ mass) reproduces the case of $\gamma\gamma$ production examined in
ref.~\cite{DelDuca:2017twk}~\footnote{It is not immediately obvious
that the limit of a massless $W$ boson should reproduce the photon
process. However, this is ultimately related to the fact that
parity-violating terms in the squared amplitude for $W$+jet production
turn out to be absent. Furthermore, the longitudinal polarisation
state does not contribute due to QED Ward identities.}.

\subsection{A momentum shift formula for $Wg$ production}
\label{sec:momshift}

Having obtained the gluonic (next-to)-soft contributions to the NLO
$Wg$ matrix element, let us now see how these can be obtained from a
momentum-shift formula analogous to those presented in
refs.~\cite{DelDuca:2017twk,vanBeekveld:2019prq}. Schematically, we
can write the contribution of a next-to-soft gluon emission from a
given Born amplitude as
\begin{equation}
  {\cal M}^{a(1)}_{\rm NLP}=\sum_{i}g_s {\bf T}^a_i
  \epsilon^\dag_\mu(k)\left[ \frac{p_{i}^\mu}{\eta_i p_i\cdot
      k+i\varepsilon} -\frac{i k^\nu J^{(i)}_{\nu\mu}} {\eta_i
      p_i\cdot k+i\varepsilon}\right]\otimes {\cal M}^{(0)}(\{p_i\}).
  \label{NLPform}
\end{equation}
Here the sum is over all external parton legs in the Born amplitude,
${\bf T}_i^a$ is a colour generator on line $i$, where we have adopted
the Catani-Seymour notation of
refs.~\cite{Catani:1997vz,Catani:1996jh}, and $\eta=\mp 1$ for an
incoming or outgoing particle respectively. There is a polarisation
vector $\epsilon^\dag(k)$ for the outgoing (next-to-soft) gluon, and
we also introduced the total angular momentum generator for each
parton leg, which can be further decomposed into its respective spin
and orbital contributions as
\begin{equation}
  J^{(i)}_{\nu\mu}=\Sigma^{(i)}_{\nu\mu}+L^{(i)}_{\nu\mu}.
  \label{Jdef}
\end{equation}
We have used the symbol $\otimes$ in eq.~(\ref{NLPform}) to mean that
the various terms must be sandwiched (where necessary) between the
external wavefunction of line $i$, and the non-radiative amplitude
${\cal M}^{(0)}(\{p_i\})$. Examples can be found throughout
refs.~\cite{DelDuca:2017twk,vanBeekveld:2019prq}, and we will see how
this works in detail below. Equation~(\ref{NLPform}) follows from the
classic works of
refs.~\cite{Low:1958sn,Burnett:1967km,DelDuca:1990gz}\footnote{For
off-shell gluons, there is an additional term in eq.~(\ref{NLPform})
that is $\propto k^\mu$. It vanishes for on-shell gluons, however,
after contraction with the polarisation vector.}. In more modern
literature on scattering amplitudes, it is known as the {\it
  next-to-soft theorem}~\cite{Cachazo:2014fwa,Casali:2014xpa} (see
e.g. refs.~\cite{White:2014qia,White:2022wbr} for details of how
things are related), and has led to the discovery of interesting
mathematical ideas relating bulk spacetime physics to a conformal
field theory living on the celestial sphere at null
infinity~\cite{Pasterski:2016qvg,Pasterski:2017kqt}. Here, we will be
much more applied, and show how eq.~(\ref{NLPform}) leads to a simple
formula for next-to-soft gluon emission, whose physical interpretation
can be elucidated further.

Let us now consider the explicit case of $Wg$ production. In applying
eq.~(\ref{NLPform}), we must start with the non-radiative amplitude
whose Feynman diagrams are given in figure~\ref{fig:LOdiags}. It will
be convenient to write the (gauge-dependent) sub-amplitude ${\cal
  M}_X$ corresponding to a given Feynman diagram $X$, where the latter
spans the labels in figures~\ref{fig:LOdiags}--\ref{fig:NLOdiags2}. In
book-keeping all possible next-to-soft contributions, we will then
follow refs.~\cite{DelDuca:2017twk,vanBeekveld:2019prq} in separating
the three different kinds of effect appearing in eqs.~(\ref{NLPform},
\ref{Jdef}).

\subsubsection{Scalar terms}
\label{sec:scalar}

The first term in the square brackets in eq.~(\ref{NLPform}) acts
multiplicatively on the whole non-radiative amplitude, with no
additional spin structure. The physics of this is that this term
corresponds to leading soft level, and hence cannot be sensitive to the
spin (or orbital angular momentum) of a given hard particle. Acting on
the non-radiative amplitude, one finds that the scalar terms sum to
\begin{equation}
  {\cal M}_{\rm scal.}=g_s\epsilon_{\mu}^\dag(k)
  \left(\frac{p_1^\mu}{-p_1\cdot k}{\bf T}_1^a
  +\frac{p_2^\mu}{-p_2\cdot k}{\bf T}_2^a
  +\frac{p_4^\mu}{p_4\cdot k}{\bf T}_4^a\right)({\cal M}_a+{\cal M}_b).
  \label{Mscal}
\end{equation}
Upon squaring the amplitude, we may evaluate all colour factors using
the relations
\begin{align}
  {\cal C}\left[{\bf T}_1^a {\cal M}_a\right]&= {\cal C}\left[{\bf
      T}_1^a {\cal M}_b\right]=C_{(A)}=C_{(D)};\notag\\
  {\cal C}\left[{\bf T}_2^a {\cal M}_a\right]&= {\cal C}\left[{\bf
      T}_2^a {\cal M}_b\right]=C_{(C)}=C_{(F)};\notag\\
  {\cal C}\left[{\bf T}_4^a {\cal M}_a\right]&= {\cal C}\left[{\bf
      T}_4^a {\cal M}_b\right]=C_{(G)}=C_{(H)},
  \label{colfacs}
\end{align}
where ${\cal C}[\ldots]$ denotes taking the colour factor of a given
diagram, and we have recognised the colour factors $\{C_{(X)}\}$ of
specific diagrams appearing in figures~\ref{fig:NLOdiags}
and~\ref{fig:NLOdiags2}. Then, evaluating all colour traces in the
squared amplitude before summing / averaging over final / initial
colours yields
\begin{equation}
  \overline{|{\cal M}_{\rm scal}|^2}=\left[C_F\left(C_F-\frac{C_A}{2}\right)
    \frac{2p_1\cdot p_2}{p_1\cdot k\,p_2\cdot k}
    +\frac{C_A^2 C_F}{2}\left(\frac{2p_1\cdot p_4}{p_1\cdot k\,p_4\cdot k}
    +\frac{2p_2\cdot p_4}{p_1\cdot k\,p_4\cdot k}\right)\right]
  \overline{|{\cal M}^{(0)}|^2}.
  \label{Mscalres}
\end{equation}
Here, we recognise the usual form of leading soft corrections to a
squared amplitude, where the individual terms that appear correspond
to separate pairs of colour charges that are linked by soft gluon
emission. For each pair or {\it dipole}, there is an appropriate
colour factor, plus a kinematic prefactor that results upon combining
the eikonal Feynman rules for the gluon. As is
well-known~\cite{Ellis:1991qj,Campbell:2017hsr}, this kinematic factor
leads to a pronounced radiation pattern, including the angular
ordering property described in the introduction, and that we will see
in more detail in section~\ref{sec:ordering}. For now, we simply note
that the remaining (next-to)soft gluon corrections will lead to
corrections to this simple radiation pattern, and our next task is to
write them in a manageable way.

\subsubsection{Spin terms}
\label{sec:spin}

Given that a coupling of the emitted gluon to the spin angular
momentum of a given hard particle is already next-to-soft level, for
the spin contributions to the total squared matrix element, we need
only worry about the interference contribution
\begin{displaymath}
  2{\rm Re}\left[{\cal M}_{\rm scal.}{\cal M}^\dag_{\rm spin}\right]
\end{displaymath}
up to next-to-leading power (NLP), where ${\cal M}_{\rm spin}$
collects all the spin effects at amplitude level. To find the latter,
we need the explicit forms of the Lorentz generators associated with
different parton legs. These are
\begin{equation}
  \Sigma^{\alpha\beta}_{ab}=\frac{i}{4}\left[\gamma^\alpha,\gamma^\beta\right]_{ab}
  \label{Sfermion}
\end{equation}
and
\begin{equation}
  \Sigma^{\alpha\beta}_{\mu\nu}=i\left(\delta^\alpha_\mu\delta^\beta_\nu-
  \delta^\alpha_\nu\delta^\beta_\mu\right)
  \label{Sgluon}
\end{equation}
for a spin-1/2 and spin-1 particle respectively, where lower indices
in these equations are spin-indices that must be contracted along the
line (n.b. lower-case Latin letters denote spinor indices). From
eq.~(\ref{NLPform}), one then finds that the spin contribution to the
NLO amplitude up to NLP level is given by
\begin{align}
  {\cal M}_{\rm spin}&=g_s\bar{v}(p_2)\left\{\frac{{\bf T}^a_1}{4p_1\cdot k}
  M^{(0)\rho\gamma}[\gamma^\mu,\slsh{k}]-\frac{{\bf T}_2^a}{4p_1\cdot k}
  [\slsh{k},\gamma^\mu]M^{(0)\rho\gamma}\right.\notag\\
  &\left.\quad
  +\frac{{\bf T}_4^a}{p_4\cdot k}
  M^{(0)\tau\gamma}\left[\delta^\mu_\tau\delta^\nu_\rho-\delta^\mu_\rho
    \delta^\nu_\tau\right]k_\nu\right\}
  u(p_1)\epsilon_\mu^\dag(k)\epsilon^\dag_\rho(p_4)\epsilon^\dag_\gamma(p_3),
  \label{Mspinres}
\end{align}
where $M^{(0)}_\sigma$ represents the non-radiative amplitude,
stripped of external wavefunctions. This must then be combined with the
scalar amplitude of eq.~(\ref{Mscalres}) and summed over polarisations
and colours to find the interference contribution. Colour factors may
again be evaluated using eq.~(\ref{colfacs}), and one may also
simplify the result by repeated use of anticommutation relations for
Dirac matrices. One finally obtains an interference contribution
\begin{equation}
  2{\rm Re}\left[{\cal M}_{\rm spin}{\cal M}^\dag_{\rm scal.}\right]
  =-2N_c C_F^2\frac{(p_1+p_2)\cdot k}{p_1\cdot k\,p_2\cdot k}
  |{\cal M}^{(0)}|^2.
  \label{spinscal}
\end{equation}

\subsubsection{Orbital angular momentum terms}
\label{sec:orbital}

For the orbital angular momentum contributions, we need the explicit
form of the angular momentum operator associated with leg $i$ in
momentum space:
\begin{equation}
  L^{\alpha\beta}=x_i^\alpha p_i^\beta-x_i^\beta p_i^\alpha=
  i\left(p_i^\alpha\frac{\partial}{\partial p_{1\beta}}-
  p_i^\beta\frac{\partial}{\partial p_{1\alpha}}\right). 
\label{Lform}
\end{equation}
Then, similarly to the spin case, the orbital angular momentum
contribution can be written as
\begin{align}
  {\cal M}_{\rm orb.}&=ig_s  \bar{v}(p_2)\left[
    \frac{{\bf T}^a_1}{p_1\cdot k}\left(
    k\cdot p_1\frac{\partial}{\partial p_{1,\mu}}-p_1^\mu k\cdot
      \frac{\partial}{\partial p_1}
      \right)M^{(0)\rho\gamma}\right.\notag\\
      &\left.\quad
      +\frac{{\bf T}^a_2}{p_2\cdot k}\left(
      p_2\cdot k\frac{\partial}{\partial p_{2,\mu}}-p_2^\mu
      k\cdot \frac{\partial}{\partial p_2}
      \right)M^{(0)\rho\gamma}\right.\notag\\
      &\left.\quad
      +\frac{{\bf T}^a_4}{p_4\cdot k}\left(
      p_4\cdot k\frac{\partial}{\partial p_{4,\mu}}-p_4^\mu
      k\cdot \frac{\partial}{\partial p_4}
      \right)M^{(0)\rho\gamma}
      \right]u(p_1)\epsilon_\mu^\dag(k)\epsilon_\rho^\dag(p_4)
  \epsilon^\dag_\gamma(p_3).
\label{Morb}
\end{align}
Once again, this must be combined with the scalar amplitude of
eq.~(\ref{Mscalres}) in order to find the relevant interference
term. Applying similar steps to those outlined in
ref.~\cite{vanBeekveld:2019prq}, we find
\begin{align}
&  2{\rm Re}\left[{\cal M}_{\rm orb.}{\cal M}^\dag_{\rm scal}\right]=
  \left\{N_cC_F^2\frac{2p_1\cdot p_2}{p_1\cdot k\,p_2\cdot k}\left[ \left(\delta
    p_{1,2}\cdot\frac{\partial}{\partial p_1} +\delta
    p_{2,1}\cdot\frac{\partial}{\partial p_2}\right) {\rm
      Tr}\left(\slsh{p}_2 M^{(0)\rho\gamma_1}\slsh{p}_1
    M^{(0)\dag}_{\rho\gamma_2}\right) \right.\right.\notag\\
    &\left.-{\rm Tr}
    \left(\delta\slashed{p}_{2,1}M^{(0)\rho\gamma_1}\slsh{p}_1
    M^{(0)}_{\rho\gamma_2}\right)
    -{\rm Tr}\left(\slashed{p}_{2}M^{(0)\rho\gamma_1}\delta\slashed{p}_{1,2}
    M^{(0)}_{\rho\gamma_2}\right)\right]\notag\\
  &+\quad\frac{C_A C_F}{2}\left[-\frac{2p_1\cdot p_2}{p_1\cdot k\, p_2\cdot k}
  \left(\slashed{p}_{1,2}\cdot\frac{\partial}{\partial p_1}+
  \slashed{p}_{2,1}\cdot\frac{\partial}{\partial p_2}
  \right)+\frac{2p_1\cdot p_4}{p_1\cdot k\, p_4\cdot k}
  \left(\slashed{p}_{1,4}\cdot\frac{\partial}{\partial p_1}-
  \slashed{p}_{4,1}\cdot\frac{\partial}{\partial p_4}
  \right)\right.\notag\\
  &\left.\left.\quad+
  \frac{2p_2\cdot p_4}{p_2\cdot k\, p_4\cdot k}
  \left(\slashed{p}_{2,4}\cdot\frac{\partial}{\partial p_2}-
  \slashed{p}_{4,2}\cdot\frac{\partial}{\partial p_4}
  \right)
  \right]{\rm Tr}\left[\slsh{p}_2M^{(0)\rho\gamma_1}\slsh{p}_1
    M^{(0)\dag}_{\rho\gamma_2}\right]\right\}{P^{\gamma_2}}_{\gamma_1},
\label{Morb2}
\end{align}
where ${P^{\gamma_2}}_{\gamma_1}$ denotes the W boson polarisation
sum. We have used the chain rule where necessary, and also introduced
the momentum shifts
\begin{equation}
  \delta p^\alpha_{i,j}=-\frac12\left(k^\alpha+\frac{p_j\cdot k}{p_i\cdot p_j}
  p_i^\alpha-\frac{p_i\cdot k}{p_i\cdot p_j}p_j^\alpha\right).
  \label{pshift}
\end{equation}
Equation~(\ref{Morb2}) looks cumbersome, but we have yet to combine it
with the spin contribution of eq.~(\ref{spinscal}). To do so, we may
first introduce a Sudakov decomposition for the emitted gluon
momentum:
\begin{equation}
  k^\mu=\frac{p_1\cdot k}{p_1\cdot p_2}p_2^\mu
  +\frac{p_2\cdot k}{p_1\cdot p_2}p_1^\mu+k_T^\mu,
  \label{Sudakov}
\end{equation}
where
\begin{equation}
  k_T\cdot p_1=k_T\cdot p_2=0.
  \label{kTperp}
\end{equation}
Equation~(\ref{pshift}) then implies
\begin{equation}
  \delta\slashed{p}_{1,2}=-\frac12\left(\slsh{k}_T
  +2\frac{p_2\cdot k}
  {p_1\cdot p_2}\slsh{p}_1\right),\quad
  \delta\slashed{p}_{2,1}=-\frac12\left(\slsh{k}_T+2\frac{p_1\cdot k}
                {p_1\cdot p_2}\slsh{p}_2\right),
  \label{Sudshifts}
\end{equation}
where upon substituting this into the second line of
eq.~(\ref{Morb2}), we may ignore the terms $\sim{\cal O}(k^\mu_T)$:
upon carrying out all Dirac traces, $k^\mu_T$ will only ever be
contracted with hard momenta in the process, to first order in soft
momentum, and all such contractions vanish. We then find
\begin{align}
  &N_c C_F^2\frac{2p_1\cdot p_2}{p_1\cdot k\,p_2\cdot k}\left[-{\rm Tr}
    \left(\delta\slashed{p}_{2,1}M^{(0)\rho\gamma_1}
    \slsh{p}_1M^{(0)}_{\rho\gamma_2}\right)
    -{\rm Tr}\left(\slashed{p}_{2}M^{(0)\rho\gamma_1}\delta\slashed{p}_{1,2}
    M^{(0)}_{\rho\gamma_2}\right)\right]{P^{\gamma_2}}_{\gamma_1}\notag\\
  &=2\frac{(p_1+p_2)\cdot k}{p_1\cdot k\,
        p_2\cdot k}|{\cal M}^{(0)}|^2.
      \label{trterms}
\end{align}
Thus, upon combining eq.~(\ref{Morb2}) with eq.~(\ref{spinscal}), the
second line of eq.~(\ref{Morb2}) is cancelled. Up to next-to-soft
level, one may then absorb the various momentum shift terms into a
redefinition of the squared nonradiative amplitude, to give the final
result
\begin{align}
  &\overline{|{\cal M}_{\rm NLP}|^2}=C_F\left(C_F-\frac{C_A}{2}\right)
  \frac{2p_1\cdot p_2}{p_1\cdot k\,p_2\cdot k}
  \overline{|{\cal M}^{(0)}|^2}(p_1+\delta p_{1,2},p_2+\delta p_{2,1},p_3,p_4)
  \notag\\
  &\quad+\frac{C_A C_F}{2}
    \frac{2p_1\cdot p_4}{p_1\cdot k\,p_4\cdot k}
    \overline{|{\cal M}^{(0)}|^2}
    (p_1+\delta p_{1,4},p_2,p_3,p_4-\delta p_{4,1})
    \notag\\
    &\quad+\frac{C_A C_F}{2}
    \frac{2p_2\cdot p_4}{p_2\cdot k\,p_4\cdot k}
    \overline{|{\cal M}^{(0)}|^2}
    (p_1,p_2+\delta p_{2,4},p_3,p_4-\delta p_{4,2}).
\label{MNLP}
\end{align}
This is our final result for the NLP matrix element, and it agrees
with a similar formula derived for prompt photon production in
ref.~\cite{vanBeekveld:2019prq}, thus showing that this is more
general than previously thought. Given eq.~(\ref{LOamp2}), we may
implement the momentum shifts as in eqs.~(\ref{pshift}, \ref{MNLP}),
and then expand to next-to-leading power using a similar method to
that outlined in the previous section. Explicit results for the three
shift terms appearing in eq.~(\ref{MNLP}) are respectively as follows:
\begin{align}
  &\overline{|{\cal M}_{\rm NLP}|^2}\Big|_1= \frac{4 g_s^4 g_w^2}{t_ku_k}
  \frac{C_F}{N_c}\left(C_F-\frac{C_A}{2}\right)
 \left[
-\frac{s (\rho^2 - 1)^2 \cos^2\theta_1 + (\rho^2 + 
       1)^2}{ (\rho^2 - 1)^2 (\cos^2\theta_1 - 
  1)} \right.\notag\\
&\left.\quad+ \frac{4(\rho - 1) (\rho + 1) (\rho^4 + 
     1) \sin\theta_1 \cos\theta_1 \cos\theta_2 \sqrt{t_k u_k}
     - (\rho^2 + 1) \rho^2 (1-\cos^2\theta_1) (t_k + 
          u_k) }{ (\rho^2 - 1)^3 (\cos^2\theta_1 - 
      1)^2}
\right];\notag\\
 &\overline{|{\cal M}_{\rm NLP}|^2}\Big|_2= \frac{2 g_s^4 g_w^2C_FC_A}{N_c}
 \Big[\frac{s ((\rho^2 - 1)^2 \cos^2\theta_1 + (\rho^2 + 
      1)^2)}{ (\rho^2 - 1)^2 t_k (\cos\theta_1 + 
    1) (-2 \sin\theta_1 \cos\theta_2 \sqrt{t_k u_k} + 
    \cos\theta_1 (t_k - u_k) + t_k + 
    u_k)} \notag\\
   &\quad+ \Big(4 \cos^2\theta_1 (\rho^2 (2 \rho^4 t_k^2 + 
          t_k^2 + \rho^2 (t_k + u_k)^2 + u_k^2) - 
       \sin\theta_1 \cos\theta_2 \sqrt{
         t_k u_k} (-2 \rho^6 (u_k - 2 t_k) + \rho^4 (t_k + 3 u_k)\notag\\
       &\quad- t_k + 
          2 \rho^2 u_k + u_k)) + 
    \cos\theta_1 ((3 \rho^6 - \rho^4 + 5 \rho^2 + 1) t_k^2 + 
       4 \sin\theta_1 \cos\theta_2 (2 (\rho - 1) (\rho + 
       1) (2 \rho^4 + 1) t_k u_k\notag\\
       &\quad\times\sin\theta_1 \cos\theta_2 +
           \sqrt{t_k u_k} (-2 \rho^6 (2 t_k + u_k) + \rho^4 (3 t_k + 
                u_k) - 2 \rho^2 (t_k + u_k) + t_k + u_k)) + (\rho - 
                1) (\rho + 1)\notag\\
                &\quad\times (7 \rho^4 + 6 \rho^2 + 3) t_k u_k + 
       4 \rho^2 (\rho^2 + 1) u_k^2) + 
    4 \rho^2 \cos^3\theta_1 ((\rho^4 + \rho^2) t_k^2 - \
(\rho^4 + 2 \rho^2 - 3) t_k u_k - (\rho^2 + 
    1) u_k^2)\notag\\
    &\quad+ (\rho^2 - 
       1)^3 \cos^4\theta_1 (\sin\theta_1 \cos\theta_2 \sqrt{t_k u_k} (u_k - 3 t_k) + t_k (t_k + u_k)) - (\rho^2 + 
       1) (\sin\theta_1 \cos\theta_2 \notag\\
       &\quad\times(8 \rho^2 t_k u_k \sin\theta_1 \cos\theta_2 - 
          \sqrt{t_k u_k} (\rho^4 (3 t_k - u_k) + 
             4 \rho^2 (t_k + 3 u_k) - 3 t_k + u_k)) + (t_k + 
             u_k)\notag\\
             &\quad\times((\rho^4 - 1) t_k + 4 \rho^2 u_k)) + (\rho^2 - 
          1)^3 t_k \cos^5\theta_1 (t_k - 3 u_k)\Big)\notag\\
       &\quad\times   \frac{1}{ (\rho^2 - 
      1)^3 t_k (\cos\theta_1 - 1) (\cos\theta_1 + 
      1)^2 (-2 \sin\theta_1 \cos\theta_2 \sqrt{t_k u_k} + 
      \cos\theta_1 (t_k - u_k) + t_k + u_k)^2}\Big];\notag\\
 &\overline{|{\cal M}_{\rm NLP}|^2}\Big|_3= \frac{2 g_s^4 g_w^2C_FC_A}
 {N_c}
\Big[-\frac{s ((\rho^2 - 1)^2 \cos^2\theta_1 + (\rho^2 + 
       1)^2}{ (\rho^2 - 1)^2 u_k (\cos\theta_1 - 
     1) (-2 \sin\theta_1 \cos\theta_2 \sqrt{t_k u_k} + 
     \cos\theta_1 (t_k - u_k) + t_k + 
     u_k)}\notag\\
  &\quad+ \Big(-4 \cos^3\theta_1 (\rho^4 (t_k^2 - 2 t_k u_k - 
           u_k^2) + (\rho^2 - 
           1)^3 u_k \sin\theta_1 \cos\theta_2 (\sqrt{t_k u_k} - 
           t_k \sin\theta_1 \cos\theta_2) \notag\\
           &\quad+ \rho^6 u_k (t_k - 
           u_k) + \rho^2 t_k (t_k + u_k)) - 
    4 \cos^2\theta_1 (\rho^2 (t_k^2 + \rho^2 (t_k + u_k)^2 + 
    2 \rho^4 u_k^2 + u_k^2) \notag\\
    &\quad+  \sin\theta_1 \cos\theta_2 ((\rho^2 - 
             1)^3 t_k u_k \sin\theta_1 \cos\theta_2 - 
          \sqrt{t_k u_k} (\rho^4 t_k + \rho^2 (4 t_k + u_k) - t_k + 
          3 \rho^6 u_k))) \notag\\
          &\quad+ 
    \cos\theta_1 (4 (\rho^4 + \rho^2) t_k^2 + 
       4 \sin\theta_1 \cos\theta_2 ((\rho^6 + \rho^4 - 
       3 \rho^2 + 1) t_k u_k \sin\theta_1 \cos\theta_2\notag\\
      &\quad -           \sqrt{t_k u_k} (\rho^4 t_k + t_k + 
             3 \rho^6 u_k - \rho^2 u_k)) + (\rho^2 - 
          1)^2 (3 \rho^2 + 1) t_k u_k + (3 \rho^6 - \rho^4 + 
          5 \rho^2 + 1) u_k^2)\notag\\
          &\quad- (\rho^2 - 
       1)^3 \cos^4\theta_1 (\sin\theta_1 \cos\theta_2 (t_k - 
          3 u_k) \sqrt{t_k u_k} + u_k (t_k + u_k)) + (\rho^2 + 
          1) (\sin\theta_1 \cos\theta_2 \notag\\
          &\quad\times(\sqrt{t_k u_k} (\rho^4 (t_k + u_k) - 4 \rho^2 (3 t_k + u_k) - t_k - 
             u_k) - 4 (\rho^4 - 2 \rho^2 - 
             1) t_k u_k \sin\theta_1 \cos\theta_2) + (t_k + 
             u_k)\notag\\
             &\quad\times(4 \rho^2 t_k + (\rho^4 - 1) u_k)) + (\rho^2 - 
             1)^3 u_k \cos^5\theta_1 (t_k + u_k)\Big)\notag\\
             &\quad\times\frac{1}{(\rho^2 - 
      1)^3 u_k (\cos\theta_1 - 1)^2 (\cos\theta_1 + 
      1) (-2 \sin\theta_1 \cos\theta_2 \sqrt{t_k u_k} + 
      \cos\theta_1 (t_k - u_k) + t_k + u_k)^2}\Big].         
\end{align}
Upon adding these results and simplifying, we find precise agreement
with the truncated NLO squared amplitude of eq.~(\ref{M1sq}), thus
confirming the validity of eq.~(\ref{MNLP}).

Comparing eq.~(\ref{MNLP}) with eq.~(\ref{Mscalres}), we see that the
effect of the next-to-soft corrections is to modify the leading power
soft gluon squared amplitude by shifting the momenta of the
nonradiative amplitude. Crucially, however, the corrections do not
modify the dipole-like nature of the result: in each term, the momenta
that are shifted in the nonradiative amplitude are the same hard
momenta that appear in the accompanying dipole radiation pattern. This
suggests a particularly nice physical interpretation of the
next-to-soft corrections, which we explore in the following section.

\section{The physics of angular-ordering breakdown}
\label{sec:ordering}

As discussed above, a well-known property of soft radiation
from pairs of (colour) charges is that it is confined to certain
cones, centered around the hard particles that emit the radiation. Put
another way, successive soft gluon emissions from the same pair of
charged particles are strongly ordered in angle, and this effect is
built into angular-ordered parton-shower algorithms to incorporate soft-gluon
interference effects in a straightforward
way~\cite{Ellis:1991qj,Campbell:2017hsr}. Given that the leading
next-to-soft gluon radiation that is captured by eq.~(\ref{MNLP})
preserves a dipole-like form, it is natural to ask whether the
momentum-shift contributions lead to a breaking or otherwise of the
angular-ordering property. We will see that, unsurprisingly, angular
ordering indeed does not persist at next-to-soft level. However, the origin of the breaking can be traced very directly to
the momentum-shift formula of eq.~(\ref{MNLP}), which allows us to understand
in physical terms how it happens.

\subsection{Angular ordering of soft radiation}
\label{sec:angsoft}

\begin{figure}
     \centering
     \begin{subfigure}[b]{0.2\textwidth}
         \centering
         \includegraphics[width=\textwidth]{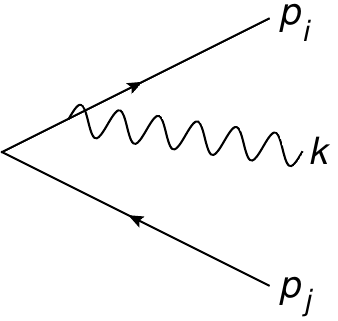}\\(a)
     \end{subfigure}
     \hspace{2cm}
     \begin{subfigure}[b]{0.2\textwidth}
         \centering
         \includegraphics[width=\textwidth]{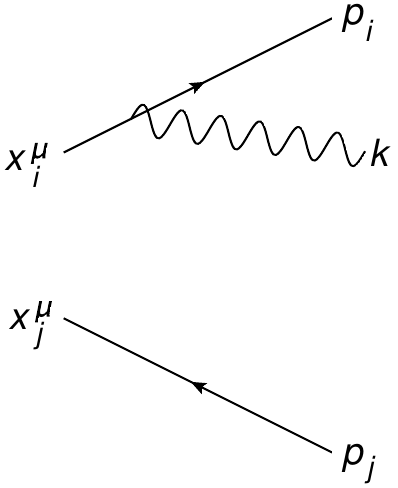}\\(b)
     \end{subfigure}
        \caption{(a) A dipole consisting of two oppositely charged
          particles, emitting a photon; (b) similar situation, taking
          into account a non-zero displacement of each fermion from
          the origin.}
        \label{fig:dipole}
\end{figure}

\begin{figure}
  \begin{center}
    \scalebox{0.6}{\includegraphics{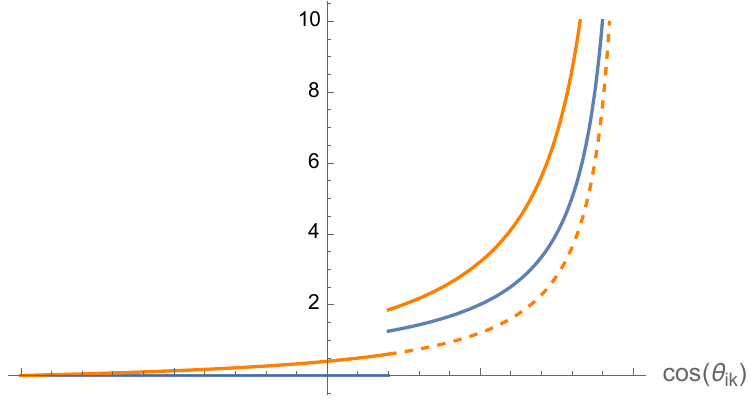}}
    \caption{Blue: distribution of soft photon radiation from a dipole
      with opening angle $\cos\theta_{ij}=0.2$; Orange: corresponding
      result coming from the leading next-to-soft correction, as
      captured by the momentum shifts in eq.~(\ref{MNLP}). The dashed
      line shows the non-angular ordered contribution to the
      next-to-soft result.}
    \label{fig:angplot}
  \end{center}
\end{figure}

Let us first recap the arguments leading to angular ordering of soft
radiation, where we will follow closely the presentation in
refs.~\cite{Ellis:1991qj,Campbell:2017hsr}. These arguments are
reproduced here to make our presentation self-contained, as
well as being necessary for the next-to-soft generalisation to be
discussed below. We will consider a final state dipole in QED,
consisting of e.g.~an electron-positron pair, as shown in
figure~\ref{fig:dipole}(a).
In the limit in which the emitted photon momentum is soft
($k^\mu\rightarrow 0$), the NLO squared amplitude for this process
assumes the form
\begin{equation}
  |{\cal M}_{\rm LP}|^2 \sim W_{ij} |{\cal M}^{(0)}|^2,
\label{MLP}
\end{equation}
where ${\cal M}^{(0)}$ is the Born amplitude, and we have introduced
the {\it radiation function}
\begin{equation}
  W_{ij}=\frac{E_k^2 p_i\cdot p_j}{p_i\cdot k\,p_j\cdot k}.
  \label{Wij}
\end{equation}
This consists of the eikonal dressing factor that we see in
e.g.~eq.~(\ref{Mscalres}), multiplied by the square of the photon
energy $E_k$ to make the radiation function dimensionless. The
energy dependence will be compensated elsewhere in the total squared
amplitude, but it is the radiation function that controls all angular
dependence of the emitted radiation. To probe the latter, it is
standard to write
\begin{equation}
  W_{ij}=W_{ij}^{[i]}+W_{ij}^{[j]},
\label{Wijsum}
\end{equation}
where the modified radiation functions appearing on the right-hand
side are given by
\begin{align}
  W_{ij}^{[i]}&=\frac12\left(W_{ij}+\frac{1}{1-\cos\theta_{ik}}
  -\frac{1}{1-\cos\theta_{jk}}\right);\notag\\
  W_{ij}^{[j]}&=\frac12\left(W_{ij}+\frac{1}{1-\cos\theta_{jk}}
  -\frac{1}{1-\cos\theta_{ik}}\right).  
\label{Wiji}
\end{align}
The reason for this -- which is not necessarily obvious a priori -- is
that the modified radiation functions have precisely the angular
ordering property noted above. That is, the soft radiation captured by
$W^{[l]}_{ij}$ is confined to a cone around particle $l$, with a
half-angle given by the angle between particles $i$ and $j$. To see
this, we can choose a Lorentz frame such that the 3-momentum
$\vec{p}_i$ is oriented along the $z$ direction, and the 3-momentum
$\vec{p}_j$ lies in the $(x,z)$ plane:
\begin{align}
  p_i^\mu&=E_i(1,0,0,1);\notag\\
  p_j^\mu&=E_j(1,\sin\theta_{ij},0,\cos\theta_{ij});\notag\\
  k^\mu&=E_k(1,\sin\theta_{ik}\cos\phi_{ik},\sin\theta_{ik}\sin\phi_{ik},
  \cos\theta_{ik}).
  \label{pijk}
\end{align}
Here $\{\theta_{ab}\}$ and $\{\phi_{ab}\}$ denote the polar and
azimuthal angles between particles $a$ and $b$ in a conventional
spherical polar coordinate system. Then
\begin{equation}
  p_j\cdot k=E_j E_k(a-b\cos\phi_{ik}),\quad
  a=1-\cos\theta_{ij}\cos\theta_{ik},\quad b=\sin\theta_{ij}\sin\theta_{ik}.
  \label{pjk1}
\end{equation}
However, by choosing an alternative frame in which the 3-momentum of
particle $j$ defines the polar axis, one may also surmise
\begin{equation}
  p_j\cdot k=E_j E_k(1-\cos\theta_{jk}),
  \label{pjk2}
\end{equation}
such that comparing eqs.~(\ref{pjk1}, \ref{pjk2}) implies
\begin{equation}
  1-\cos\theta_{jk}=a-b\cos\phi_{ik}.
  \label{angrel}
\end{equation}
The integration over the final-state phase space will include an
integral over the azimuthal angle $\phi_{ik}$ of the emitted photon,
and a useful intermediate step in integrating $W_{ij}^{[i]}$ is to
consider the integral
\begin{equation}
  I_{ij}^{[i]}=\int_0^{2\pi}\frac{d\phi_{ik}}{2\pi}\frac{1}
  {1-\cos\theta_{jk}},
  \label{Iijdef}
\end{equation}
which occurs in the first and third terms of eq.~(\ref{Wiji}), whose
explicit form in angular coordinates is
\begin{equation}
  W_{ij}^{[i]}=\frac12\left(
  \frac{1-\cos\theta_{ij}}{(1-\cos\theta_{ik})(1-\cos\theta_{jk})}
  +\frac{1}{1-\cos\theta_{ik}}-\frac{1}{1-\cos\theta_{jk}}
  \right).
  \label{Wijangles}
\end{equation}
Following refs.~\cite{Ellis:1991qj,Campbell:2017hsr}, we may transform
to $z=e^{i\phi_{iq}}$, such that eq.~(\ref{Iijdef}) becomes a contour
integral around the unit circle in the complex $z$-plane:
\begin{equation}
  I_{ij}^{[i]}=\frac{1}{i\pi b}\oint\frac{dz}{(z-z_+)(z-z_-)},\quad
  z_\pm=\frac{a}{b}\pm\sqrt{\frac{a^2}{b^2}-1}.
  \label{Iij2}
\end{equation}
Only the pole at $z=z_-$ lies inside the unit circle, such that one
may carry out the integral using Cauchy's theorem, yielding
\begin{equation}
  I_{ij}^{[i]}=\frac{1}{\sqrt{a^2-b^2}}=\frac{1}
  {|\cos\theta_{ik}-\cos\theta_{ij}|}.
  \label{Iijres}
\end{equation}
Substituting this result into eq.~(\ref{Wijangles}), one may rearrange
to give
\begin{equation}
  W_{ij}^{[i]}=\frac{1}{2(1-\cos\theta_{ik})}\left[
    1+\frac{(\cos\theta_{ik}-\cos\theta_{ij})}
    {|\cos\theta_{ik}-\cos\theta_{ij}|}
    \right].
  \label{Wijresult}
\end{equation}
A plot of this function is shown in figure~\ref{fig:angplot} in blue, for
$\cos\theta_{ij}=0.2$. We see that there is non-zero radiation only
for polar angles around particle $i$ which are less than or equal to
the opening angle $\theta_{ij}$ between the dipole, as
expected. Furthermore, the divergence at $\cos\theta_{ik}\rightarrow
1$ coincides with the emitted photon becoming collinear with particle
$i$.

Here we have explicitly considered QED, where this pronounced
radiation pattern is known as the Chudakov effect. The same arguments
apply to QCD in the case of single gluon radiation from a dipole,
where this is an overall colour
singlet~\cite{Ellis:1991qj,Campbell:2017hsr}. In both cases, the
simple quantum mechanical argument for the suppressed radiation
outside the cone is that the wavelength of the emitted photon becomes
such that it cannot resolve the individual (colour) charges in the
dipole.

\subsection{Next-to-soft radiation from a dipole}
\label{sec:dipoleNLP}

Having recalled how angular ordering arises from soft radiation, let
us now examine how the additional momentum-shift corrections in
eq.~(\ref{MNLP}) modify the picture. To this end, we may again
restrict ourselves to the simplest possible case of a final-state
dipole in gauge theory, namely the electron-positron pair of
figure~\ref{fig:dipole}(a). Then the effect of an additional photon
emission up to next-to-soft level is to modify eq.~(\ref{MLP}) to
\begin{equation}
  |{\cal M}_{\rm NLP}|^2 \sim W_{ij} |{\cal M}^{(0)}(p_i-\delta p_{i,j},
  p_j-\delta p_{j,i})|^2,
\label{MNLP2}
\end{equation}
using the momentum shift definitions of eq.~(\ref{pshift}).\footnote{An
explicit derivation of this formula can be found in
ref.~\cite{vanBeekveld:2019prq}, and uses similar steps to those
presented here in section~\ref{sec:momshift}.} Up to NLP level, we
can expand to first order in the momentum shifts, and also use the
fact that the squared Born interaction depends only on the Mandelstam
invariant $s=2p_i\cdot p_j$, to get
\begin{equation}
  |{\cal M}_{\rm NLP}|^2\sim W_{ij} f(s)
  -2\Big[\delta p_{i,j}\cdot p_j+\delta p_{j,i}\cdot p_i\Big]W_{ij}f'(s),
  \label{MNLP3}
\end{equation}
where
\begin{equation}
  f(s)=|{\cal M}^{(0)}|^2
  \label{fs}
\end{equation}
is the squared Born amplitude with unshifted kinematics, and the prime
denotes its first-order derivative. To examine
the angular properties of the next-to-soft term, we define the
dimensionless radiation functions
\begin{align}
	\widetilde{W}^{[l]}_{ij}&=-\left(\frac{\delta p_{i,j}\cdot p_j}{E_k E_j} + \frac{\delta p_{j,i}\cdot p_i}{E_k E_i}\right)W^{[l]}_{ij},
	\label{Wijtilde}
\end{align}
such that
\begin{equation}
  \widetilde{W}_{ij} = \widetilde{W}^{[i]}_{ij}+\widetilde{W}^{[j]}_{ij} =-\left(\frac{\delta p_{i,j}\cdot p_j}{E_k E_j} + \frac{\delta p_{j,i}\cdot p_i}{E_k E_i}\right)W^{[l]}_{ij}
  \label{Wtildesum}
\end{equation}
controls the total next-to-soft correction to the radiation
pattern. Using the parametrisation of eq.~(\ref{pijk}), we have
\begin{align}
  \widetilde{W}^{[i]}_{ij}=(1-\cos\theta_{ik})W_{ij}^{[i]}
+(1-\cos\theta_{jk})W_{ij}^{[i]}.  
\label{Wtildecalc1}
\end{align}
In the first term, the prefactor is independent of the azimuthal angle
$\phi_{iq}$, and thus does not affect the integration over the
latter. We can thus reuse the previous result for the azimuthal
integration of $W_{ij}^{[i]}$ when calculating the radiation
pattern. In the second term of eq.~(\ref{Wtildecalc1}), the prefactor
cancels the singularity in $\cos\theta_{jk}$, such that one finds
\begin{align}
  \int_0^{2\pi}\frac{d\phi_{ik}}{2\pi} \widetilde{W}^{[i]}_{ij}
  &=\frac12\left[1+\frac{\cos\theta_{ik}-\cos\theta_{ij}}
    {|\cos\theta_{ik}-\cos\theta_{ij}|}\right]\notag\\
  &\quad+\frac12\int_0^{2\pi}\frac{d\phi_{ik}}{2\pi}\left[
    \frac{1-\cos\theta_{ij}}{1-\cos\theta_{ik}}
    +\frac{a-b\cos\phi_{ik}}{1-\cos\theta_{ik}}-1\right]\notag\\
  &=\frac12\left[1+\frac{\cos\theta_{ik}-\cos\theta_{ij}}
    {|\cos\theta_{ik}-\cos\theta_{ij}|}\right]
  +\sin^2\left(\frac{\theta_{ij}}{2}\right)
  \cot^2\left(\frac{\theta_{ik}}{2}\right).
  \label{Wtilderes}
\end{align}
This is our final result for the azimuthally-averaged next-to-soft
radiation pattern from one particle in a dipole. Interestingly, it has
the form of a sum of an angular-ordered term, analogous to the pure
soft case, plus a breaking term, which has a remarkably compact
analytic form. We show this function in figure~\ref{fig:angplot}, and
can see clearly the effect of the angular-ordering term, in that there
is a discontinuity at $\cos\theta_{ik}=\cos\theta_{ij}$. However, for
$\cos\theta_{ik}<\cos\theta_{ij}$, corresponding to gluon emission
outside the cone region, there is indeed a non-zero radiation
distribution. For completeness, we show the effect of the
non-angular-ordered term (i.e. the second term in
eq.~(\ref{Wtilderes})) inside the cone. This is shown as the dashed
line in figure~\ref{fig:angplot}, and we see that it smoothly joins
the radiation outside the cone, as it should. There remains a (hard)
collinear singularity around particle $i$, which acts as a
next-to-soft correction to a collinear emission which is also strictly
soft. In interpreting the figure, we must remember that the LP and NLP
radiation functions ($W^{[i]}_{ij}$ and $\tilde{W}^{[i]}_{ij}$
respectively) are defined with different energy ratios to make them
dimensionless. Thus, the overall normalisation between the two curves
in figure~\ref{fig:angplot} is not particularly meaningful. Rather,
figure~\ref{fig:angplot} illustrates the significant qualitative
difference between the angular distributions of soft and next-to-soft
radiation: the latter breaks angular ordering.

\begin{figure}
  \begin{center}
    \scalebox{0.6}{\includegraphics{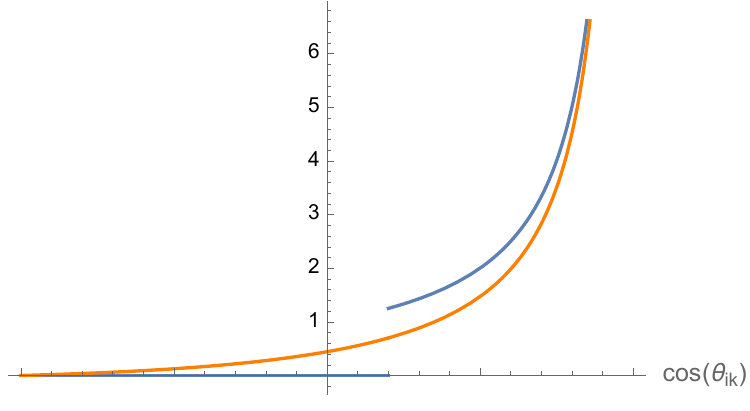}}
    \caption{Blue: distribution of soft photon radiation from a dipole
      of massless particles with opening angle $\cos\theta_{ij}=0.2$;
      Orange: corresponding result if particle $j$ is massive, such
      that $v_j=0.5$ in eq.~(\ref{massivecase}).}
    \label{fig:angplot2}
  \end{center}
\end{figure}

Note that a similar effect appears for massive emitters at leading
soft level, as is
well-known~\cite{Ellis:1991qj,Campbell:2017hsr}. Denoting the
energy-normalised velocity of the two dipole legs with $v_i$ and $v_j$
respectively, one obtains for the azimuthally-averaged emission
pattern
\begin{eqnarray}
  \int^{2\pi}_0 \frac{d \phi_{ik}}{2\pi} W_{ij}^{[i]} = \frac{v_i}{2(1-v_i \cos\theta_{ik})}\left[\frac{v_i - \cos\theta_{ik}}{1-v_i \cos\theta_{ik}} + \frac{\cos\theta_{ik} - v_j \cos\theta_{ij}}{\sqrt{(\cos\theta_{ik} - v_j \cos\theta_{ij})^2 + \sin\theta_{ik}^2(1-v_j^2)}} \right].
  \label{massivecase}
\end{eqnarray}
This reduces to the massless form when $v_i = v_j = 1$. However, when
$v_j \neq 1$ (but $v_i = 1$) the sharp transition is replaced by a
smooth damping of the total radiation from the dipole, which extends
to angles larger than the opening angle of the dipole.  This is shown
in figure~\ref{fig:angplot2}, which contrasts the massless and massive
cases. Here, the presence of an intrinsic momentum scale results in
the breaking of angular-ordering, and it would be interesting to further
examine the interplay between next-to-soft and massive effects.

Let us now understand the next-to-soft effect in more physical
terms. We can do this given that radiation outside the cone is
associated with the specific momentum shifts in eq.~(\ref{pshift}),
whose origin is the orbital angular momentum contributions to the
squared matrix element. One can only generate such an orbital angular
momentum if the two worldlines of the fermions in the dipole are
mutually displaced. In particular, let us choose the origin of
spacetime such that both lines are displaced to 4-positions $x^\mu_i$
and $x^\mu_j$, as shown in figure~\ref{fig:dipole}(b). The
relationship between such displacements and next-to-soft corrections
was considered in ref.~\cite{Laenen:2008gt}, which used Schwinger
proper time methods to write the scattering amplitude for hard
particles emitting radiation in terms of quantum mechanical
(first-quantised) path integrals over their spacetime
trajectories. More specifically, one can show that such an amplitude
is given by
\begin{equation}
  {\cal A}(p_1,\ldots, p_n)=\int{\cal D}A_\mu 
  \left[\prod_{i=1}^n \int d^d x_i e^{-ip_i\cdot x_i}f(x_i,p_i;A_\mu)
    \right]H(x_1\ldots x_n;A_\mu)e^{iS[A_\mu]},
  \label{ampsoft}
\end{equation}
where $H(x_1,\ldots x_n;A_\mu)$ is a {\it hard function} that produces
the outgoing particles at initial positions $\{x_i\}$ with final
momenta $\{p_i\}$. The formal definition of this quantity can be found
in ref.~\cite{Laenen:2008gt}, and will not be needed in what
follows. The path integral in eq.~(\ref{ampsoft}) is over the
(next-to) soft gauge field, and includes the usual dependence on the
action $S[A_\mu]$, where we suppress the dependence on the matter
fields for brevity. Associated with each hard particle is an integral
over its initial position $x_i$, a certain exponential factor, and
further factor $f(x_i,p_i;A_\mu)$, which for scalar particles is
\begin{equation}
  f(x_i,p_i;A_\mu)=\int{\cal D}z_i
  \exp\left[i\int_0^\infty d\tau\left(\frac{\dot{z}_i^2}{2}
    +q_i(\beta+\dot{x})\cdot A(x_i+\beta_i\tau+z_i)+\frac{iq_i}{2}
    \partial\cdot A(x_i+\beta_i \tau+z_i)\right)
    \right].
  \label{fdef}
\end{equation}
In this equation, we have parametrised the spacetime trajectory of the
$i^{\rm th}$ particle via
\begin{equation}
  x^{(i)}(t)=x_i+\beta_i\tau+z_i(\tau),
  \label{xitau}
\end{equation}
where $\beta_i$ is the 4-velocity associated with the final momentum
$p_i$, and $\tau$ the proper time. The quantity $z_i(\tau)$ then
constitutes a fluctuation about the classical straight-line
trajectory, and the path integral over $z_i(\tau)$ corresponds to
summing over all possible fluctuations. This path integral is carried
out subject to the boundary conditions of fixed initial position $x_i$
and final momentum $p_i$ for each particle. Finally, $q_i$ is the
electric charge of hard particle $i$.

As shown in ref.~\cite{Laenen:2008gt}, the path integral over
worldline trajectories in eq.~(\ref{xitau}) can be carried out
perturbatively. The leading term -- corresponding to keeping the
classical trajectory only -- amounts to the hard particle not
recoiling, and thus emitting pure soft radiation only. Expanding
order-by-order in $z_i(\tau)$ then amounts to including all possible
wobbles in the spacetime trajectory which, by the uncertainty
principle, amounts to including the emission of radiation at
progressively subleading orders in the momentum of the emitted
radiation. By keeping the first-order term only,
ref.~\cite{Laenen:2008gt} found a set of Feynman rules corresponding
to the emission of next-to-soft radiation from hard
particles. Repeating the analysis for fermionic emitting particles
leads to an extra term in eq.~(\ref{fdef}), that corresponds to the
spin-dependent part of the next-to-soft theorem of
eqs.~(\ref{NLPform}, \ref{Jdef}).

Now let us focus on the contribution to eq.~(\ref{ampsoft}) that stems
from the initial separation of the dipole members, namely the non-zero
initial positions $\{x_i\}$. At next-to-soft level, keeping track of
these non-zero positions means that the path integral in
eq.~(\ref{fdef}) can be replaced by its leading term (i.e. the
classical trajectory only). The hard particle factor of
eq.~(\ref{fdef}) then reduces to the well-known Wilson line describing
the emission of soft radiation~\cite{Laenen:2008gt}:
\begin{align}
  f(x_i,p_i;A_\mu)&\rightarrow \exp\left[iq_i\int_0^\infty d\tau
    \beta_i^\mu A_\mu(x_i+\beta_i \tau)\right]\notag\\
  &=1-q_i\int\frac{d^d k}{(2\pi)^d}\frac{p_i^\mu \tilde{A}_\mu(k)}
  {p_i\cdot k}e^{ip_i\cdot k}+{\cal O}(q_i^2),
  \label{flimit}
\end{align}
where we have transformed to momentum space in the second line, and
expanded in the coupling so as to isolate the effect of a single
photon emission in the second term. Next, we can expand the
exponential appearing in the $k$ integral, where the first subleading
correction corresponds to the next-to-soft contribution. Collecting
these factors on all lines, the effect of the non-zero initial
positions in the path integral of eq.~(\ref{ampsoft}) is
\begin{align}
&  \left[\prod_{i=1}^n \int d^d x_i e^{-ix_i\cdot p_i}
    \right]\left[
    \sum_{j=1}^n q_j \frac{d^d k}{(2\pi)^d} (-ix_j\cdot k)\frac{p_j\cdot A(k)}
    {p_j\cdot k}
    \right]H(p_1,\ldots,p_n)=\Gamma^\mu A_\mu,
  \label{position1}
\end{align}
where carrying out the integrals over the positions $\{x_i\}$ yields
\begin{equation}
  \Gamma^\mu=\int\frac{d^d k}{(2\pi)^d}\sum_{j=1}^n q_j\left(
  \frac{p_j^\mu}{p_j\cdot k} k_\nu\frac{\partial}{\partial p_{j\nu}}
  \right)H(p_1,\ldots,p_n).
  \label{position2}
\end{equation}
In the path integral over the gauge field, this looks like an
additional Feynman rule for the emission of a single photon from each
line, which involves a derivative acting on the hard function. In
fact, the result of eq.~(\ref{position1}) is incomplete. In
eq.~(\ref{ampsoft}), the hard function depends upon the gauge field,
as it must. Expanding this order-by-order in the coupling amounts to
including the effects of soft gluon emissions from inside the hard
interaction (see ref.~\cite{Laenen:2008gt} for a detailed explanation). As
shown in the very early work of ref.~\cite{Low:1958sn}, such
contributions can be fixed by gauge invariance. The most
straightforward way to implement this here is to note that the factor
$\Gamma^\mu$ will form part of a complete scattering amplitude ${\cal
  A}^\mu$ for the emission of a photon of momentum $k$, which must
satisfy the Ward identity $k_\mu {\cal A}^\mu=0$. This in turn implies
that we must modify
\begin{equation}
  \Gamma^\mu\rightarrow \tilde{\Gamma}^\mu,\quad
  k_\mu\tilde{\Gamma}^\mu=0.
  \label{Ward}
\end{equation}
Requiring a local combination of momenta yields the unique result (see
also ref.~\cite{Bern:2014oka})
\begin{equation}
  \tilde{\Gamma}^\mu=
  \int\frac{d^d k}{(2\pi)^d}\sum_{j=1}^n q_jk_\nu\left(
  \frac{p_j^\mu}{p_j\cdot k} \frac{\partial}{\partial p_{j\nu}}
  -  \frac{p_j^\nu}{p_j\cdot k} \frac{\partial}{\partial p_{j\mu}}
  \right)H(p_1,\ldots,p_n).
  \label{position3}
\end{equation}
Comparison with eq.~(\ref{Lform}) allows to explicitly recognise the
form of the orbital angular momentum of each hard particle, a fact
which was not clarified in ref.~\cite{Laenen:2008gt}. However, it
makes precise the above expectation, that non-zero initial positions
of the dipole members will indeed give rise to the orbital angular
momentum part of the next-to-soft theorem. 

The physics of angular-ordering breaking is then as follows. Soft
radiation has an infinite Compton wavelength, and thus is unable to
see the separation between the two fermion worldlines, as they emanate
from a given hard interaction. Next-to-soft radiation, on the other
hand, is able to resolve the length scale corresponding to the
difference in initial particle positions, which manifests itself in
the orbital angular momentum contributions being non-zero, as captured
by the momentum shifts in eq.~(\ref{MNLP}). The fact that wide-angle
radiation now sees the initial ``size'' of the dipole means that it
will no longer see a zero net charge. Hence, radiation can be present
outside the cone.


\section{Conclusion}
\label{sec:conclude}

In this paper, we have performed a case study looking at the physical
interpretation of next-to-soft radiation. The characterisation of such
radiation is of great interest in furthering the precision frontier at
current collider experiments, as well as addressing interesting
conceptual questions in field theory. In addition to building up new
methods and techniques, it is important to build intuition about
next-to-soft physics, that may in turn inform further developments.
With this motivation in mind, we have here focused on a particular
formula for incorporating gluon radiation using a dipole-like formula
that incorporates next-to-soft effects through shifts of the momenta
appearing in the non-radiative amplitude. This formula first appeared
for colour-singlet final states in ref.~\cite{DelDuca:2017twk}, and
was extended to particular processes with partons in the final state
in ref.~\cite{vanBeekveld:2019prq}. We have here checked its validity
in another process ($Wg$ production). Next, we looked at
the physical consequences of this formula, which stem from the fact
that it has the form of a sum of dipole-like contributions.

One of the most well-known properties of soft emission from dipoles is
that interference effects lead to suppression of the radiation outside
cones surrounding each hard particle, whose half-angle corresponds to
the opening angle between the constituents of the dipole. Faced with
eq.~(\ref{MNLP}), then, we can ask if the inclusion of next-to-soft
corrections breaks the angular-ordering property. Indeed it does, and
the physical mechanism of this is that the momentum shifts capture
precisely that part of the next-to-soft physics -- orbital angular
momentum contributions -- that is associated with an initial
separation between the dipole constituents. This provides a new length
scale, which the radiation is then able to resolve. Although this
creates some radiation outside the cones described above, there still
remains a significant discontinuity in the radiation distribution at
the edge of the cone.

We hope that our results provide useful physical intuition to
researchers working in this area, as well as inspiring further similar
studies.

\section*{Acknowledgments}
EL and AT would like to thank the MHRD Government of India, for the
SPARC grant SPARC/2018-2019/P578/SL, entitled {\it Perturbative QCD
  for Precision Physics at the LHC.}  CDW is supported by the UK
Science and Technology Facilities Council (STFC) grants ST/P000258/1
and ST/T000759/1. MvB was supported by a Royal Society Research
Professorship (RP$\backslash$R1$\backslash$180112), and by the STFC
grant ST/T000864/1. SP would like to thank Satyajit Seth for useful
discussions.  SP would like to thank the Physical Research Laboratory,
Department of Space, Govt.~of India, for a Post Doctoral
Fellowship. AD would like to thank CSIR, Govt. of India, for an SRF
fellowship (09/1001(00 49)/2019- EMR-I).

\bibliographystyle{JHEP}
\bibliography{refs}

\end{document}